\documentclass[pra,twocolumn,preprintnumbers,amsmath,amssymb,showpacs,nofootinbib,floatfix]{revtex4}

\usepackage{graphicx,bm}

\makeatletter
\def\graphicscale{\twocolumn@sw{0.3}{0.4}}
\def\graphicthreescale{\twocolumn@sw{0.3}{0.4}}

\begin{document}

\title{Three-dimensional antiferromagnetic CP$^{N-1}$ models }

\author{Francesco Delfino$^1$, Andrea Pelissetto$^2$, and Ettore Vicari$^1$} 

\address{$^1$ Dipartimento di Fisica dell'Universit\`a di Pisa
        and INFN, Largo Pontecorvo 3, I-56127 Pisa, Italy}
\address{$^2$ Dipartimento di Fisica dell'Universit\`a di Roma ``La Sapienza"
        and INFN, Sezione di Roma I, I-00185 Roma, Italy}

\date{\today}

\begin{abstract}

We investigate the critical behavior of three-dimensional
antiferromagnetic CP$^{N-1}$ (ACP$^{N-1}$) models in cubic lattices,
which are characterized by a global U($N$) symmetry and a local U(1)
gauge symmetry. Assuming that critical fluctuations are associated
with a staggered gauge-invariant (hermitian traceless matrix) order
parameter, we determine the corresponding Landau-Ginzburg-Wilson (LGW)
model. For $N=3$ this mapping allows us to conclude that the
three-component ACP$^2$ model undergoes a continuous transition that
belongs to the O(8) vector universality class, with an effective
enlargement of the symmetry at the critical point.  This prediction is
confirmed by a detailed numerical comparison of finite-size data for
the ACP$^2$ and the O(8) vector models.  We also present a
renormalization-group (RG) analysis of the LGW theories for $N\ge
4$. We compute perturbative series in two different renormalization
schemes and analyze the corresponding RG flow.  We do not find stable
fixed points that can be associated with continuous transitions.

\end{abstract}


\maketitle



\section{Introduction}
\label{intro}

CP$^{N-1}$ models are a class of models in which the fundamental field
is a complex $N$-component unit vector (more precisely, an element of
the complex projective manifold CP$^{N-1}$), and which are
characterized by a global U($N$) symmetry and a local U(1) gauge
symmetry.  They emerge as effective theories of SU($N$) quantum
antiferromagnets~\cite{RS-90,Kaul-12,KS-12,BMK-13} and of scalar
electrodynamics with a compact U(1) gauge group.  The simplest
three-dimensional (3D) CP$^{N-1}$ lattice model is defined by the
Hamiltonian
\begin{equation}
H = J \sum_{\langle ij \rangle} | \bar{\bm{z}}_i \cdot {\bm z}_j |^2,
\label{hcpn}
\end{equation}
where the sum is over the nearest-neighbor sites of a cubic lattice,
${\bm z}_i$ are $N$-component complex vectors satisfying
$\bar{\bm{z}}_i\cdot {\bm z}_i=1$. The model is ferromagnetic for
$J<0$, antiferromagnetic for $J > 0$.

The CP$^1$ model can be mapped onto the O(3)-symmetric Heisenberg
model. Indeed, if one defines O(3) spins $s_i^\alpha = \sum_{ab}
\bar{z}_i^a \sigma_{ab}^\alpha z_i^b$, where $\sigma^\alpha$ are the
Pauli matrices, one can rewrite the CP$^1$ Hamiltonian as that of the
usual 3-vector Heisenberg model. As a consequence, the critical
properties can be straightforwardly derived by using the wealth of
results available for the Heisenberg model, see, e.g.,
Refs.~\cite{PV-02,CHPRV-02,HV-11}. On the other hand, several aspects
of the phase behavior of CP$^{N-1}$ models with $N>2$ remain unclear
and worth being further investigated.

The critical behavior of these models can be investigated by
constructing an effective Landau-Ginzburg-Wilson (LGW) theory.  This
approach requires the identification of the order parameter associated
with the critical modes. Assuming that the critical behavior is
essentially driven by gauge-invariant modes, a plausible choice for
ferromagnetic ($J<0$) systems is the gauge-invariant site
variable~\cite{NCSOS-13,Kaul-12}
\begin{equation}
Q_{i}^{ab} = \bar{z}_i^a z_i^b - {1\over N} \delta^{ab},
\label{qdef}
\end{equation}
which is a hermitian and traceless $N\times N$ matrix.  In the
corresponding LGW theory, the fundamental field is therefore the most
general traceless hermitian matrix $\Phi^{ab}(x)$, which one can
imagine being defined as the average of $Q_{i}^{ab}$ over a large but
finite lattice domain. The Hamiltonian is the most general
fourth-order polynomial in $\Phi$ consistent with the U($N$) symmetry:
\begin{eqnarray}
{\cal H} &=& {\rm Tr} (\partial_\mu \Phi)^2 
+ r \,{\rm Tr} \,\Phi^2 +  w \,{\rm tr} \,\Phi^3 \label{hlg}\\
&& +  \,u\, ({\rm Tr} \,\Phi^2)^2  + v\, {\rm Tr}\, \Phi^4 .
\nonumber
\end{eqnarray}
For $N=2$ one recovers the O(3)-symmetric LGW theory because the cubic
term vanishes and the two quartic terms are equivalent
\cite{footnote-Phi}, consistently with the equivalence between the
CP$^1$ and the Heisenberg model.  Because of the presence of the cubic
term, on the basis of mean-field arguments, one expects the system to
undergo a first-order transition for any $N>2$, unless the Hamiltonian
parameters are tuned so that $w = 0$ in the effective model.  This
prediction is, however, contradicted by recent numerical studies
\cite{NCSOS-11,KHI-11,Kaul-12,NCSOS-13}, which find evidence of
continuous transitions in models that are expected to be in the same
universality class as that of the 3D CP$^2$ model.  In particular, a
numerical study of 3D loop models~\cite{NCSOS-13} provided the
estimate $\nu=0.536(13)$ for the correlation-length critical
exponent. These results imply the existence of a 3D CP$^2$
universality class characterized by a U(3) global symmetry and U(1)
gauge invariance, with a corresponding fixed point (FP) that cannot be
determined in perturbation theory at fixed $N$. In order to access
this FP, the authors of Ref.~\cite{NCSOS-13} proposed a
double expansion around $N=2$ (where the cubic term vanishes) and
$\epsilon=4-d$, arguing that a continuous transition may be possible
for values of $N$ sufficiently close to $N=2$. For larger values of
$N$, i.e. $N\ge 4$, the numerical
analyses~\cite{NCSOS-13,NCSOS-11,Kaul-12} show evidence of first-order
transitions.  In the large-$N$ limit, the quantum field theory
corresponding to the ferromagnetic CP$^{N-1}$ model gives rise to an
effective abelian Higgs model and Landau-Ginzburg (LG) theory of
superconductivity~\cite{MZ-03}, whose renormalization-group (RG) flow
presents a stable FP for a sufficiently large number of
components. Thus continuous transitions are again possible in
CP$^{N-1}$ models at large $N$, sharing the same universal critical
behaviors of the LG theory of superconductivity~\cite{MZ-03}.  These
results are again in contrast with the conclusions obtained from the
LGW theory (\ref{hlg}) and suggest that, at least for large values of
$N$, critical modes are not exclusively associated with the
gauge-invariant order parameter $Q$, cf. Eq.~(\ref{qdef}).

In this paper we investigate the critical behavior of
antiferromagnetic CP$^{N-1}$ (ACP$^{N-1}$) models, such as those
described by the Hamiltonian (\ref{hcpn}) with $J>0$, on a cubic
lattice (we expect a similar behavior on any bipartite lattice).  For
$N=2$, they undergo a critical transition in the same universality
class as that of the ferromagnetic CP$^1$ model.  Indeed, the ACP$^1$
model is equivalent to the antiferromagnetic Heisenberg model, which
in turn can be mapped onto the ferromagnetic one by performing the
transformation ${\bm s}(x) \to (-1)^{x_1 + x_2 + x_3} {\bm s}(x)$,
where $x \equiv (x_1,x_2,x_3)$. Therefore, the staggered variables ${\bm
  s}_{\rm stag} = (-1)^{x_1 + x_2 + x_3} {\bm s}$ or $Q_{\rm stag} =
(-1)^{x_1 + x_2 + x_3} Q$ have the same critical behavior as ${\bm s}$
or $Q$ in the ferromagnetic model.  However, for $N > 2$ the behavior
of ACP$^{N-1}$ models differs from that of ferromagnetic CP$^{N-1}$
models, as we shall show.

Keeping the assumption that the critical modes can be represented by
local staggered gauge-invariant variables, we show that the LGW
Hamiltonian describing the behavior of the critical modes in the
ACP$^{N-1}$ models is the one given in Eq.~(\ref{hlg}), without the
cubic term, that is with $w = 0$.  Indeed, the staggered nature of the
order parameter gives rise to a symmetry $\Phi\to -\Phi$, which
prevents the presence of odd terms in $\Phi$, such as the cubic term.
This fact greatly simplifies the RG analysis of the theory.  In
particular, it allows us to predict that the critical behavior of the
ACP$^2$ model belongs to the universality class of the O(8) vector
model, with a dynamical enlargement of the symmetry at the critical
point. Correspondingly, we predict $\nu\approx 0.85$ and $\eta\approx
0.03$ for the ACP$^2$ model, which differ from those found in
Refs.~\cite{NCSOS-11,NCSOS-13} for the ferromagnetic CP$^2$
universality class.  To validate this prediction, we perform Monte
Carlo (MC) simulations of the lattice ACP$^2$ and O(8) vector
models. The critical exponents and finite-size scaling (FSS) functions
turn out to be the same in the two models, in agreement with the RG
argument.  Finally, we present a general RG study of the LGW theory
(\ref{hlg}) without cubic term. We compute high-order
field-theoretical (FT) perturbative series in two different schemes.
The RG analysis does not provide evidence of the existence of stable
FP for $N\ge 4$.

We mention that a critical-point symmetry enlargement analogous to
that of ACP$^2$ models occurs in the antiferromagnetic RP$^{2}$ model
~\cite{FMSTV-05,ACFJMRT-05}. These systems are similar to those
considered here: their Hamiltonian is also given by Eq.~(\ref{hcpn}),
but the site variable is a {\em real} unit vector. In the RP$^{2}$
case, an analogous RG argument based on the corresponding LGW theory
shows that the model should have a critical behavior in the
universality class of the O(5) vector model.

The paper is organized as follows. In Sec.~\ref{lgw} we construct the
LGW theory which is expected to describe the critical modes at
continuous transitions of ACP$^{N-1}$ models, assuming a staggered
gauge-invariant order parameter.  Sec.~\ref{acp2} is devoted to a
numerical study of the ACP$^2$ model. We show that its continuous
transition is in the same universality class as that of the O(8)
vector model.  In Sec.~\ref{hopa} we study the RG flow relevant for
models with more components, i.e., for $N\ge 4$, computing and
analyzing high-order FT perturbative series for the corresponding LGW
theories.  Finally in Sec.~\ref{conclu} we summarize our main results
and draw some conclusions. Some details are reported in the
appendices.

\section{LGW theory for the ACP$^{N-1}$ models}
\label{lgw}

We now derive the LGW theory for the critical modes at the transition
in ACP$^{N-1}$ models, which is completely specified by the symmetry
of the model, the nature of the order parameter, and the
symmetry-breaking pattern.  In the case at hand, the model has a
global U($N$) symmetry and a local U(1) gauge invariance.  We assume
that the critical modes are effectively represented by local
gauge-invariant variables, such as (\ref{qdef}).  In the case of
antiferromagnetic interactions ($J>0$), the minimum of the Hamiltonian
(\ref{hcpn}) is locally realized by taking $\bar{\bm z}_i\cdot {\bm
  z}_j = 0$ for any pair of nearest-neighbor sites.

In order to construct the LGW Hamiltonian, we should identify the
order parameter of the transition. At variance with the ferromagnetic
case, in the ACP$^{N-1}$ model we should take into account the
explicit breaking of translational invariance in the low-temperature
phase.  To clarify the issue, let us consider the antiferromagnetic
O($M$) vector model with Hamiltonian $H_{\rm O} = \sum_{\langle
  ij\rangle} {\bm s}_i \cdot {\bm s}_j$.  In this case the order
parameter is ${\bm \phi} = \sum_x p_x {\bm s}_x$, where $p_x$ is the
parity of the site $x \equiv (x_1,x_2,x_3)$ defined by $p_x =
(-1)^{\sum_i x_i}$. Under translations of one site, we find ${\bm
  \phi} \to -{\bm \phi}$, hence this parameter allows us to probe the
breaking of translational invariance. In the ACP$^{N-1}$ model the
natural field variable is the combination $\bar{z}_i^a z_i^b$, which
is invariant under the local U(1) gauge transformations of the
model. Hence, we define the order parameter
\begin{equation}
B^{ab} = \sum_{\langle x \rangle} p_x \bar{z}^a_x z^b_x.
\label{biab}
\end{equation}
It is immediate to verify that $B$ is hermitian and traceless [this
  follows from the presence of $p_x$], that it changes sign under
translations of one site which exchange the two sublattices, and that
it coincides with the O(3) order parameter for $N = 2$.  Then, as
usual, in order to construct the LGW model, we replace $B$ with a
local variable $\Psi$ as fundamental variable (essentially, one may
imagine that $\Psi$ is defined as $B$, but now the summation extends
only over a large, but finite, cubic sublattice).  Then, we write down
the most general fourth-order polynomial that is invariant under
U($N$) transformations and under the ${\mathbb Z}_2$ transformation
$\Psi \to -\Psi$, a consequence of the translation invariance of the
original theory. We obtain
\begin{eqnarray}
{\cal H}_a = {\rm Tr} (\partial_\mu \Psi)^2 
+ r \,{\rm Tr} \,\Psi^2 +
{u_0\over 4} \, ({\rm Tr} \,\Psi^2)^2  + {v_0 \over 4} \, {\rm Tr}\, \Psi^4 .
\label{ahlg} 
\end{eqnarray}
The original U($N$) symmetry corresponds to the symmetry $\Psi \to
U^\dagger \Psi U$ where $U\in {\rm U}(N)$.  The order parameter $\Psi$
is a hermitean traceless matrix as the variable $\Phi$ introduced in
the ferromagnetic case.  However, because of the presence of the
symmetry $\Psi\to-\Psi$, which is a specific feature of the
antiferromagnetic model, the LGW Hamiltonian (\ref{ahlg}) does not
present a cubic term, which instead appears in the $\Phi^4$ Hamiltonian
(\ref{hlg}) corresponding to the ferromagnetic case.

Note that the model is not only characterized by the symmetry group, 
but also by the nature of the order parameter. There are indeed other 
models with U($N$) symmetry (see, e.g., Ref.~\cite{BPV-05} for an example),
which, however, have a different order parameter and different
symmetry-breaking patterns, leading to different universality classes.

The stability domain of ${\cal H}_a$ can be determined by studying
the asymptotic behavior of the potential
\begin{equation}  
V(\Psi) =  r \,{\rm Tr} \,\Psi^2 +
{u_0 \over 4} \, ({\rm Tr} \,\Psi^2)^2  + {v_0\over 4} \, {\rm Tr}\, \Psi^4 .
\label{vpsi}
\end{equation}
This analysis can be easily performed by noting that $V(\Psi)$ only
depends on the $N$ real eigenvalues $\lambda_a$ of the hermitian
matrix $\Psi$, which satisfy the condition $\sum_a \lambda_a = 0$. We
find that the theory is stable if
\begin{eqnarray}
u_0 + b_N v_0 > 0, \qquad
b_N = {N^2 - 3 N  + 3\over N(N-1) } ,
\label{stabcond1}
\end{eqnarray}
and  if
\begin{eqnarray}
&u_0 + {1\over N} v_0 > 0  \quad & {\rm for}\; {\rm even }\;N,\label{stabcond2}\\
&u_0 + c_N v_0 > 0  \quad & {\rm for}\; {\rm odd }\;N,\nonumber 
\end{eqnarray}
where 
\begin{equation}
c_N = {N^2 + 3\over N(N^2-1)}.
\label{stabcond3}
\end{equation}
Physical systems corresponding to the effective theory (\ref{ahlg})
with $u_0,v_0$ that do not satisfy these constraints are expected to
undergo a first-order phase transition.

The analysis of the minima of the potential $V(\Psi)$ for $r<0$
gives us information on the symmetry-breaking patterns. For $v_0<0$,
the absolute minimum of $V(\Psi)$ is realized by configurations with
$\Psi = U\Psi_{\rm min} U^\dagger$ and
\begin{equation}
\Psi_{\rm min} \sim
\left( 
\begin{array}{l@{\ \ }l@{\ \ }}
I_{N-1}& \quad 0 \\
\;\;0 & -(N-1) \\
\end{array}
\right), 
\label{psivp}
\end{equation}
where $I_{n}$ indicates the $n\times n$ identity matrix.  This gives
rise to the symmetry-breaking pattern
\begin{equation}
{\rm U}(N) \to {\rm U}(1)\times {\rm U}(N-1).
\label{vnpat}
\end{equation}
On the other hand, for $v_0>0$ and even $N$ the minumum is 
realized by
\begin{equation}
\Psi_{\rm min} \sim
\left( 
\begin{array}{l@{\ \ }l@{\ \ }}
I_{N/2}& \;\; 0 \\
\;\; 0 & -I_{N/2} \\
\end{array}
\right)
\label{psivn}
\end{equation}
implying  the symmetry-breaking pattern
\begin{equation}
{\rm U}(N) \to {\rm U}(N/2)\times {\rm U}(N/2).
\label{vppat}
\end{equation}
For $v_0>0$ and odd values of $N$, we have instead
\begin{equation}
\Psi_{\rm min} \sim
\left( 
\begin{array}{l@{\ \ }l@{\ \ }}
I_{(N+1)/2}& \;\; 0 \\
\;\; 0 & - k I_{(N-1)/2} \\
\end{array}
\right),
\label{psivn2}
\end{equation}
$k = (N+1)/(N-1)$, so that 
\begin{equation}
{\rm U}(N) \to {\rm U}(N/2+1/2)\times {\rm U}(N/2-1/2).
\label{vppat2}
\end{equation}
Note that for $N=3$, the symmetry breaking patterns (\ref{vnpat}) and 
(\ref{vppat2}) are equivalent, hence the sign of $v_0$ does not play any role.

An important remark is in order. The derivation of the LGW Hamiltonian
(\ref{ahlg}) is based on the assumption that the order parameter is
the staggered and traceless hermitian matrix (\ref{biab}). This
assumption can be checked for $N = 3$. As we show in
App.~\ref{appmin}, the minimum-energy configurations of Hamiltonian
(\ref{hcpn}) for $J>0$ have a very simple structure. Modulo a global
U(3) transformation, one can take ${\bm z}_i = (1,0,0)$ on one
sublattice, and ${\bm z}_i = (0,a_i,b_i)$ on the other one. Hence, a
zero-temperature configuration corresponds to
\begin{equation}
B = \begin{pmatrix}
1 & 0 & 0 \\ 0 & -\sum_i |a_i|^2 & -\sum_i a^*_i b_i \\ 0 & 
                 -\sum_i a_i b_i^* & - \sum_i |b_i|^2 \end{pmatrix}.
\end{equation}
Therefore, $B$ is nonvanishing in the low-temperature phase and 
represents the correct order parameter.
The symmetry-breaking pattern is therefore that given in Eq.~(\ref{vnpat})
or, equivalently, Eq.~(\ref{vppat2}).

For $N\ge 4$, we have not been able to identify ordered
zero-temperature configurations that are translation invariant at
least on one sublattice, hence, we have not been able to check that
$B$, as defined in Eq.~(\ref{biab}), is nonvanishing in the
low-temperature phase, hence that it can be taken as the order
parameter.  In the following, we make the working hypothesis that this
is the case, determining what this assumption implies for the nature
of the transitions in the ACP$^{N-1}$ models.

For $N=3$ the LGW theory (\ref{ahlg}) simplifies.
Indeed, one can easily prove that
\begin{eqnarray}
{\rm Tr}\, \Psi^4 = {1\over 2} ({\rm Tr} \,\Psi^2)^2  
\label{quar}
\end{eqnarray}
for any $3\times3$ traceless hermitean matrix.  Then, let us define an
8-component real vector field $\phi$ as follows:
\begin{eqnarray}
&& {\Psi_{11} + a_+ \Psi_{22}\over \sqrt{2}} = \phi_1,\quad 
{\Psi_{11} + a_- \Psi_{22}\over \sqrt{2}} = \phi_2,
\label{idphps} \\
&& \Psi_{12} = \phi_3 + i \phi_4,\quad
\Psi_{13} = \phi_5 + i \phi_6,\quad
\Psi_{23} = \phi_7 + i \phi_8 \nonumber ,
\end{eqnarray}
where $a_\pm = (1 \pm\sqrt{3})/2$. In terms of the field $\phi$ we have
\begin{equation}
{1\over 2} {\rm Tr}\,\Psi^2 = \phi \cdot \phi.
\label{phcorr}
\end{equation}
The, we can rewrite ${\cal H}_a$ as 
\begin{eqnarray}
{\cal H}_{{\rm O}} = (\partial_\mu \phi)^2 + 2 r \,{\phi}^2 + 
{g_0} \, (\phi^2)^2, 
\label{hlgv} 
\end{eqnarray}
where $g_0 \equiv u_0+v_0/2$, proving that the model is equivalent to 
the O(8) vector theory.

Since the O(8) vector theory has a stable Wilson-Fisher FP, the above
correspondence allows us to predict that the critical behavior of the
ACP$^2$ model must share the same universal features as the 3D O(8)
vector model.  Note that the enlargement of the symmetry to O(8) is a
feature of the LGW theory (\ref{ahlg}), i.e., of the expansion up to
fourth powers of $\Psi$. Indeed, one can easily check that the sixth-order 
terms allowed by the U(3) symmetry, such as ${\rm Tr}\,\Psi^6$,
do not share the O(8) symmetry.  Since these terms are RG irrelevant
at the O(8) FP, the contribution of the terms breaking the O(8)
symmetry is suppressed at the critical point.  In this sense
we have a dynamic enlargement of the symmetry to O(8) at the critical
point.

Of course, the fact that only one quartic term is independent for
$N=3$ holds also for the Hamiltonian (\ref{hlg}) corresponding to the
ferromagnetic CP$^2$ model.  Thus, it has an O(8) FP, which is
unstable due to the presence of the cubic term. This cubic term gives
rise to a spin-3 perturbation of the O(8) model~\cite{CPV-03}. Its RG
dimension $y_3$ can be estimated using the results of
Ref.~\cite{DPV-03} for the crossover exponent $\phi_3$,
i.e. $y_3=\phi_3/\nu$. The crossover exponent $\phi_3$ at the O(8) FP
was estimated in Ref.~\cite{DPV-03}, obtaining $\phi_3=0.97(3)$ in the
fixed-dimension massive-zero-momentum scheme, and $\phi_3=0.95(5)$ in
the $\epsilon$ expansion.  Since $\nu= 0.85(2)$ (see Sec.~\ref{mco8}),
we obtain $y_3\approx 1.1$. As $y_3 > 0$, the O(8) FP is unstable in
the presence of the cubic term, and, therefore, the O(8) FP cannot be
the stable FP in the case of ferromagnetic interactions.  This is
confirmed by the results of Refs.~\cite{NCSOS-11,NCSOS-13}. The
critical exponents, for instance $\nu=0.536(13)$, significantly differ
from those of the O(8) universality class.

It is worth mentioning that an analogous enlargement of the symmetry
at the critical point is also observed~\cite{FMSTV-05,ACFJMRT-05} in
the antiferromagnetic RP$^{2}$ model.  The Hamiltonian of RP$^{N-1}$
models is analogous to that of CP$^{N-1}$ models It is given by
Eq.~(\ref{hcpn}), with real $N$-component spins ${\bm s}_i$ replacing
the complex vectors ${\bm z}_i$.  The order parameter should be a
symmetric and traceless $N\times N$ matrix $\Sigma$, analogous to the
matrix $B$ defined in Eq.~(\ref{biab}). The corresponding LGW
Hamiltonian is given in Eq.~(\ref{ahlg}), with $\Psi$ replaced by
$\Sigma$ \cite{FMSTV-05,ACFJMRT-05}.  For $N=3$, the two quartic terms
are proportional, and one obtains the $\Phi^4$ theory of the O(5)
vector model (again high-order terms break the O(5) symmetry, but
since they are irrelevant, the O(5) symmetry breaking is suppressed in
the critical limit).  The numerical results of
Refs.~\cite{FMSTV-05,ACFJMRT-05} for the antiferromagnetic RP$^2$
model confirm this prediction.

To identify the nature of a critical transition for $N\ge 4$, if it
exists, we must determine the fixed points of the RG flow of the LGW
theory (\ref{ahlg}).  The absence of a stable FP implies the absence
of continuous transitions. Such an analysis is quite complex, due to
the presence of two quartic terms. We will perform it in
Sec.~\ref{hopa}.

\section{Critical behavior of the ACP$^2$ lattice model}
\label{acp2}

We now check the predictions of the previous section for the
critical behavior of the 3D ACP$^2$ model, confirming that it
undergoes a continuous transition in
the O(8) universality class.

\subsection{Monte Carlo simulations and observables}
\label{MCsim}

In order to study the critical behavior of the ACP$^2$ lattice model
(\ref{hcpn}) with $J=1$, we perform Monte Carlo simulations 
of cubic systems of linear size $L$ with
periodic boundary conditions. Because of the antiferromagnetic nature 
of the model we take $L$ even. We use a standard Metropolis algorithm, 
hence we are only able to obtain reliable results up to 
$L = 40$. We use a simple updating algorithm. If 
$\varphi = ({\rm Re}\, z, {\rm Im}\, z)$ is a six-component vector,
the update consists in proposing the new vector $R\varphi$, where 
$R$ is a random O(2) matrix acting on two randomly chosen components
of $\varphi$.

\begin{figure}[tbp]
\includegraphics*[scale=\graphicscale]{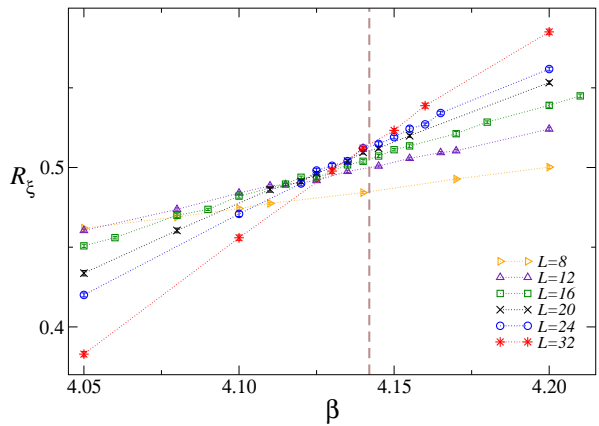}
\caption{(Color online) MC data of $R_\xi$ for the ACP$^2$ lattice
  model for different lattice sizes $L$.  They show a crossing point
  at $\beta\approx 4.14$.  The dotted lines are drawn to guide the
  eye.  The vertical dashed line corresponds to our best estimate
  (\ref{betacacp}) of $\beta_c$ obtained by a FSS analysis of the
  available data assuming the transition to belong to the O(8)
  universality class. }
\label{rxi}
\end{figure}

In our MC simulations we compute correlations of the gauge invariant
operator
\begin{equation}
P_{{x}}^{ab} = \bar{z}_{x}^a z_{x}^b. 
\label{piab}
\end{equation}
Its two-point correlation function is defined as
\begin{equation}
G({x}-{y}) = \langle {\rm Tr}\, P_{x}^\dagger  
P_{y} \rangle = 
  \langle |\bar{\bm z}_{x} \cdot {\bm z}_{y}|^2 \rangle.
\label{gxyp}
\end{equation}
Due to the staggered nature of the ordered parameter, we should 
distinguish correlations between points belonging to the same 
sublattice and points belonging to different sublattices. Here 
we define the susceptibility and the correlation length by 
summing only over points with the same parity:
\begin{eqnarray}
&&\chi =  \sum_{x\; {\rm even} } G({x}) = \widetilde{G}({0}), 
\label{chisusc}\\
&&\xi^2 \equiv  {1\over 4 \sin^2 (p_{\rm min}/2)} 
{\widetilde{G}({0}) - \widetilde{G}({p})\over 
\widetilde{G}({p})},
\label{xidefpb}
\end{eqnarray}
where ${x}$ runs over all even points, 
\begin{equation}
\widetilde{G}({p})=\sum_{{x} \; {\rm even}} e^{ip\cdot x} G({x})
\label{widedef}
\end{equation}
is the Fourier transform of $G({x})$ over the even
sublattice, ${p} = (p_{\rm min},0,0)$, and $p_{\rm min} \equiv 2
\pi/L$. Finally, we consider the Binder parameter 
\begin{equation}
U = { \langle [\sum_{x\; {\rm even}} {\rm Tr}\, P_{0}^\dagger  
P_{x} ]^2 \rangle \over  
\langle \sum_{{x}\; {\rm even}} {\rm Tr}\, P_{0}^\dagger  P_{x} \rangle^2 } .
\label{binderdef}
\end{equation}
In the FSS limit any RG invariant quantity $R$, such as
$R_\xi\equiv \xi/L$ and $U$, is expected to behave as
\begin{eqnarray}
R(\beta,L) = f_R( X ) + L^{-\omega} g_R(X) + \ldots
\label{rsca}
\end{eqnarray}
where $X = (\beta-\beta_c)L^{1/\nu}$ and $f_R(X)$ is a universal
function apart from a trivial normalization of the argument. In
particular, the quantity $R^* \equiv f_R(0)$ is universal within the
given universality class.  The approach to the asymptotic behavior is
controlled by the universal exponent $\omega>0$, which is associated
with the leading irrelevant RG operator.  Around $\beta_c$ one may
expand $f_R(X)$ and $g_R(X)$ in powers of the scaling variable $X$,
obtaining
\begin{equation}
R = R^* + \sum_{n=1} b_n (\beta-\beta_c)^n L^{n/\nu} + 
L^{-\omega} \sum_{n=0} c_n (\beta-\beta_c)^n L^{n/\nu} + ...
\label{fitr}
\end{equation}
The exponent $\eta$ is determined by analyzing the FSS behavior of 
the susceptibility 
\begin{equation}
\chi \sim L^{2-\eta} \left[ f_\chi(X) + O(L^{-\omega})\right].
\label{chisca}
\end{equation}
We present a FSS analysis of the numerical data of the ACP$^2$ lattice
model, up to $L=40$. In Fig.~\ref{rxi} we show $R_\xi\equiv \xi/L$ for
several values of $L$. Data have a crossing point, providing evidence
of a transition in the interval $4.1\lesssim \beta \lesssim 4.2$.
An analogous behavior is shown by the Binder parameter $U$.

To determine whether the transition is continuous or of first order,
we should estimate the effective exponent $\nu$ that gives the slope
of the data at the critical point.  At a first-order transition we
expect $\nu=1/d=1/3$~\cite{NN-75,FB-82,PF-83}, while $\nu>1/d$ at
continuous transitions.  The numerical data, including those of the
specific heat, definitely exclude a first-order transition. Indeed,
the increase of $dR/d\beta$ at the crossing point is much slower than
$L^3$.  Therefore, we conclude that the ACP$^2$ model has a continuous
transition.

\subsection{The 3D O(8) vector model}
\label{mco8}

\begin{figure}[tbp]
\includegraphics*[scale=\graphicscale]{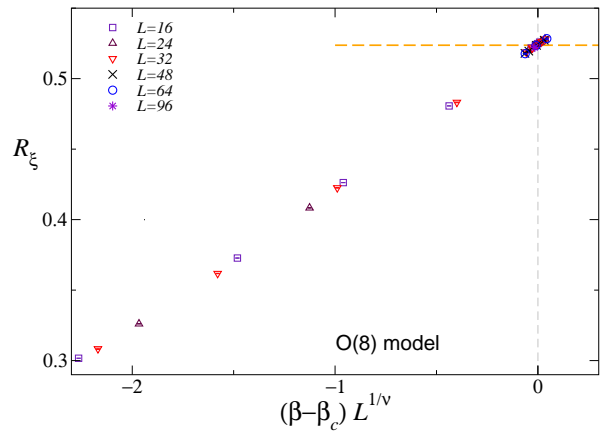}
\caption{(Color online) FSS behavior of $R_\xi$ for the
  O(8) vector model. Plot of $R_\xi$ versus $(\beta-\beta_c)L^{1/\nu}$,
  setting $\beta_c=1.92677$ and $\nu=0.85$.  The dashed horizontal line
  corresponds to $R_\xi=0.5237$.  }
\label{rxiscao8}
\end{figure}

To show that the ACP$^2$ lattice model belongs to the O(8)
universality class, we show that critical exponents and FSS curves are
the same in the two models. Thus, we begin by computing these
quantities in the O(8) spin model defined by the Hamiltonian
\begin{equation}
H_{\rm O} = -  \sum_{\langle ij \rangle} {\bm s}_i \cdot {\bm s}_j ,
\label{ho}
\end{equation}
where the spin variable ${\bm s}_i$ is an 8-component unit vector.
We consider cubic systems of linear size $L$ 
with periodic boundary conditions.

We consider the two-point
function
\begin{equation}
G_o({x}-{y}) = \langle {\bm s}_{{x}} \cdot {\bm s}_{{y}}\rangle,
\label{gxypo}
\end{equation}
and compute the corresponding susceptibility and second-moment
correlation length. They are defined as in Eqs.~(\ref{chisusc}) and
(\ref{xidefpb}), but now we sum over all lattice points, as the model
is ferromagnetic.  Moreover, we consider the Binder parameter $U$
defined as in Eq.~(\ref{binderdef}): we replace $P_{\bm x}$ with ${\bm
  s}_{x}$ and sum over all lattice points.

We perform simulations on lattices of size up to $L=96$ (we use a
cluster algorithm) and estimate $R_\xi$, $U$, and $\chi$. To compute
the critical exponents we fit the data to Eqs.~(\ref{rsca}) and
(\ref{chisca}).  In the analysis we take into account the scaling
corrections of order $L^{-\omega}$, fixing $\omega\approx 0.8$, as
predicted by the FT perturbative analyses discussed below.
Corrections turn out to be small, hence the analyses of the MC data
are unable to provide a more accurate estimate of $\omega$.  Fits of
$R_\xi$ to Eq.~(\ref{fitr}) (we take the first terms in the
expansions) give $\beta_c = 1.92677(2)$,
\begin{eqnarray}
\nu = 0.85(2), \quad \eta=0.0276(5),
\label{o8critexp}
\end{eqnarray}
and the universal critical value $R_\xi^* = 0.5237(4)$.  The quoted
uncertainty includes the statistical error and the variation of the
estimates as $\omega$ varies between 0.75 and 0.85, an interval that
is larger than that obtained in the FT analyses reported below.  In
Fig.~\ref{rxiscao8} we show $R_\xi$ versus $(\beta-\beta_c)L^{1/\nu}$
using the above-reported estimates. Data collapse onto a single curve,
confirming the accuracy of the estimates. Consistent results are
obtained from the analysis of the Binder parameter, which also gives
$U^* = 1.0383(3)$.

The values of the critical exponents can be compared with previous
results. Field theory gives $\nu\approx 0.830$ and $\eta\approx 0.027$
\cite{AS-95}, while the analysis of strong-coupling expansions gives
$\nu\approx 0.84,\,0.86$ \cite{BC-97} (the two estimates are obtained
by means of two different resummation methods). Within errors, they
agree with the estimates (\ref{o8critexp}). We have repeated the
analysis of the available six-loop series within the massive
zero-momentum renormalization scheme~\cite{BNGM-77,AS-95}, using the
conformal mapping method that exploits the known large-order behavior
of the perturbative expansions~\cite{LZ-77,GZ-98}.  We obtain
$\nu=0.826(4)$, $\eta=0.025(1)$, and $\omega=0.81(1)$, where the
errors are related to the change of the estimates with respect to a
(reasonable) variation of the parameters entering the resummation
procedure. They are substantially consistent with our favorite MC
estimates (\ref{o8critexp}), although one may suspect that errors are
slightly underestimated. The FT analysis also provides the estimate of
$\omega$ that we used (note that, to be on the safe side, we allowed
for a much larger uncertainty in the MC analysis).

\subsection{FSS of the ACP$^2$ lattice model}
\label{fssacp2}

If the transitions in the ACP$^2$ and O(8) models belong to the same
universality class, critical exponents and FSS curves $f_R(X)$ for RG
invariant quantities (apart from a trivial rescaling of the variable
$X$) should be the same.  This is what we check below.

\begin{figure}[tbp]
\includegraphics*[scale=\graphicscale]{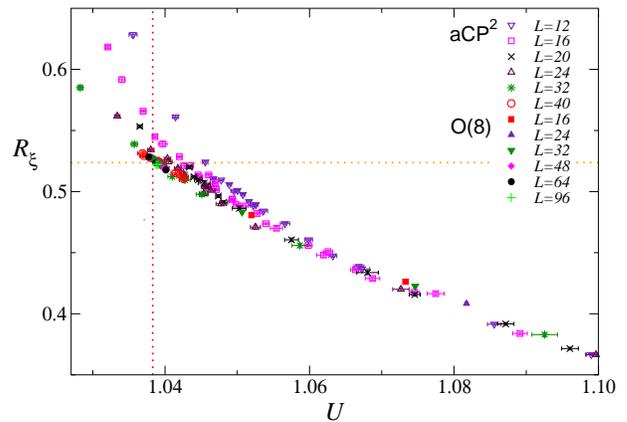}
\caption{(Color online) Plot of $R_\xi$ vs $U$. ACP$^2$ and O(8)
  results approach the same asymptotic curve. The dotted horizontal
  and vertical lines correspond to the values $R_\xi^*\approx 0.5237$
  and $U^*\approx 1.0383$ estimated in the O(8) vector model.  }
\label{rxibi}
\end{figure}

A first unbiased universality check which does not need an estimate of
the critical point $\beta_c$, is obtained by plotting $R_\xi$ versus
$U$.  Indeed, since both $R_\xi$ and $U$ satisfy Eq.~(\ref{rsca}), we
must have
\begin{equation}
R_\xi = F(U) + O(L^{-\omega}),
\label{rxibisca}
\end{equation}
where $F(U)$ is a universal function. In Fig.~\ref{rxibi} we compare
the results for the two models: they appear to approach the same
asymptotic curve with increasing the lattice size.  Scaling
corrections are consistent with the expected $L^{-\omega}$ behavior
with $\omega\approx 0.8$.  They are much smaller for $\beta<\beta_c$
than for $\beta > \beta_c$.

\begin{figure}[tbp]
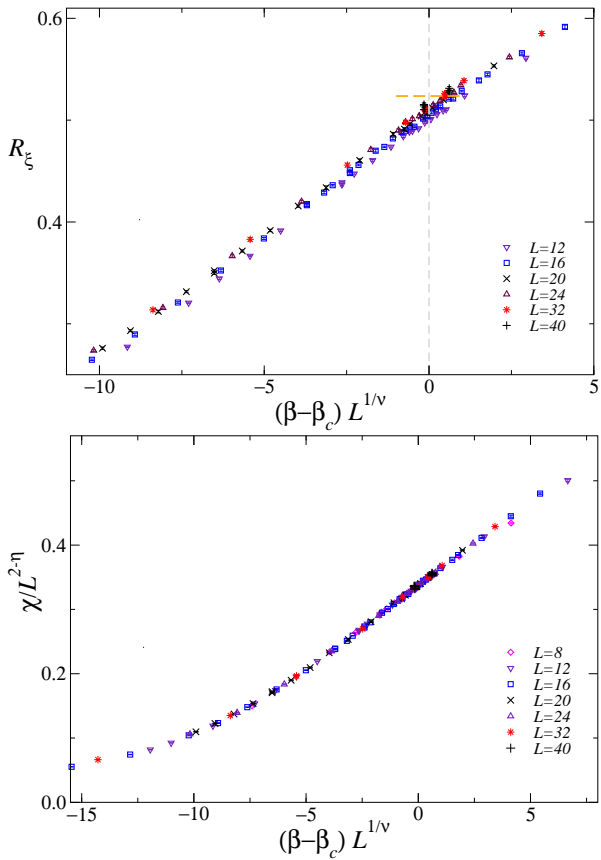

\includegraphics*[scale=\graphicscale]{rxisca.eps}
\includegraphics*[scale=\graphicscale]{chisca.eps}
\caption{(Color online) Scaling behavior of the ratios
  $\chi/L^{2-\eta}$ and $R_\xi$ vs $(\beta-\beta_c)L^{1/\nu}$ with
  $\beta_c=4.142$ and the critical exponents of the O(8) universality
  class, cf. Eq.~(\ref{o8critexp}).  In the top figure the dashed
  horizontal segment indicates the O(8) universal value
  $R_\xi=0.5237$, which is clearly approached by the data of the
  ACP$^2$ lattice model with increasing $L$.}
\label{chirxisca}
\end{figure}

The dependence of the data on the inverse temperature $\beta$ around
the crossing point is consistent with the O(8) results.  Fits of the
data around the crossing point to the first few terms of the
expansions appearing in Eq.~(\ref{fitr}) provide an accurate estimate
of the critical point,
\begin{equation}
\beta_c = 4.142(1).
\label{betacacp}
\end{equation}
In Fig.~\ref{chirxisca} we show $\chi L^{\eta-2}$ and $R_\xi$ versus
$X$, using the estimate (\ref{betacacp}) of $\beta_c$ and the O(8)
estimates (\ref{o8critexp}) of the critical exponents. The data
approach asymptotic scaling curves. Scaling corrections are larger for
$R_\xi$, but definitely compatible with the expected $L^{-\omega}$
behavior. Note also (not shown) that the scaling curves of $R_\xi$ for
the ACP$^2$ and O(8) nicely match after a trivial rescaling of the
scaling variable $(\beta-\beta_c)L^{1/\nu}$: if we define $Y = X$ for
the ACP$^2$ and $Y = 3.9 X$ for the O(8) model, all data fall on the
same curve when plotted versus $Y$.  Finally, we consider the specific
heat $C_v$ at $\beta_c$. We expect $C_v \approx a + cL^{\alpha/\nu}$
with $\alpha/\nu=2/\nu-3\approx -0.67$.  The MC data are consistent
with this behavior.

In conclusion the numerical analysis of the ACP$^2$ lattice model
provides a robust evidence that its continuous transition belongs to
the universality class of the O(8) vector model, as predicted by the
RG arguments of Sec.~\ref{lgw}.

\section{RG flow for  $N\ge 4$}
\label{hopa}

In this section we present a FT study of the RG flow of the LGW theory
(\ref{ahlg}) with $N\ge 4$, i.e. the most general $\Phi^4$ theory with
traceless hermitian $N\times N$ matrix fields and parity symmetry.
The critical behavior at a continuous transition is controlled by the
FPs of the RG flow, which are determined by the common zeroes of the
$\beta$ functions associated with the quartic parameters.  The
presence of a stable FP controls the universal features of the
critical behavior in the case of continuous transition.  The absence
of a stable FP implies the absence of a corresponding universality
class, hence a transition characterized by the same symmetry-breaking
pattern must be of first order.

\subsection{The ${\overline {\rm MS}}$ perturbative scheme}
\label{msbar}

We compute the $\beta$ functions of the quartic couplings in the
${\overline {\rm MS}}$ renormalization scheme~\cite{TV-72}, which uses
the dimensional regularization around four dimensions, and the
modified minimal-subtraction prescription.  Thus the RG functions are
obtained from the divergences for $\epsilon\equiv 4-d \to 0$ appearing
in the perturbative expansion of the correlation functions of the
critical massless theory.

The procedure is straightforward, see, e.g., Ref.~\cite{ZJ-book}.  The
renormalized couplings are defined from the irreducible four-point
correlation function, and the ${\overline {\rm MS}}$ $\beta$ functions
are
\begin{equation}
\beta_u(u,v) = \left. \mu{\partial u\over \partial \mu}\right|_{u_0,v_0},\quad
\beta_v(u,v) = \left. \mu{\partial v\over \partial \mu}\right|_{u_0,v_0},
\label{betafms}
\end{equation}
where $\mu$ is the renormalization energy scale of the $\overline{\rm
  MS}$ scheme. Here, $u$ and $v$ are the renormalized couplings
corresponding to $u_0$, $v_0$, defined so that $u\propto u_0/\mu^\epsilon$ and
$v\propto v_0/\mu^\epsilon$ at the lowest order. We compute
the $\beta$ functions up to five loops. The complete series for
$N=4$ are reported in App.~\ref{hopert}.

\subsubsection{One-loop analysis close to four dimensions}
\label{oneloop}

Let us first analyze the one-loop $\beta$ functions.  They read
\begin{eqnarray}
&&\beta_u = -\epsilon u  + {N^2+7\over 6} u^2 + {2N^2 - 3\over 3N} uv
+ {N^2 + 3\over 2 N^2 } v^2,\qquad  \label{betau}\\
&&\beta_v = -\epsilon v  + 2uv + {N^2-9\over 3N} v^2 .
\label{betav}
\end{eqnarray}
The exact normalization of the renormalized variables
can be easily read from these series.

Since for $N=2$ and $N=3$ the two quartic terms are not independent,
an appropriate combination of the above $\beta$ functions must
reproduce the $\beta$ functions of the O(3) and O(8) $\Phi^4$
theories.  Indeed, using Eq.~(\ref{quar}) and setting $g=u+v/2$, we
obtain
\begin{equation}
\beta_u + {1\over 2}\beta_v  =  \beta_{{\rm O}(3)}(g) 
= - \epsilon g + {11\over 6} g^2 
\label{p3bf}
\end{equation}
for $N=2$, and 
\begin{equation}
\beta_u + {1\over 2}\beta_v =
\beta_{{\rm O}(8)}(g) = -\epsilon g + {8\over 3} g^2
\label{p8bf}
\end{equation}
for $N=3$. These exact relations provide a stringent check of 
the five-loop series for the model (\ref{ahlg}), which must reproduce 
the corresponding series of the O(3) and O(8) vector
model~\cite{oneepsexp1,oneepsexp2} for $N=2$ and 3.

For $N\ge 4$ the FPs of the RG flow are given by the common zeros of
the $\beta$ functions (\ref{betau}) and (\ref{betav}).  Their
stability requires that the eigenvalues of the matrix $\Omega_{ij} =
\partial \beta_{g_i}/ \partial g_j$ (where $g_{1,2}$ correspond to
$u,v$) have positive real part. In the standard $\epsilon$-expansion
scheme~\cite{WF-72}, the FPs, i.e., the common zeroes of the
$\beta$-functions, are determined perturbatively as expansions in
powers of $\epsilon\equiv 4-d$, while exponents are obtained by
expanding the corresponding RG functions computed at the FP in powers
of $\epsilon$.

A straightforward analysis of the one-loop $\beta$ functions
(\ref{betau}) and (\ref{betav}) finds four different FPs.
Two of them have $v = 0$ and are always unstable.
We have the trivial Gaussian FP at $(u=0,v=0)$,
which is always unstable with respect to both quartic perturbations.
There is also an O($M$) symmetric FP with $M=N^2-1$ at
\begin{equation}
 u=\epsilon \,{6\over N^2+7},\quad v=0, 
 \label{onfp}
 \end{equation}
which can be shown, nonperturbatively, to be unstable with respect to the 
operator ${\rm Tr}\,\Psi^4$.
Indeed, such operator contains a spin-4 perturbation with respect to the
  O($M$) group~\cite{CPV-03}, which is relevant for any $M>4$ to
  $O(\epsilon)$, and for any $M\ge 3$ in three
  dimensions~\cite{CPV-00,HV-11}.  

There are also two FPs with $v <0$. One of them is stable, the other
is unstable. However, they only exist for $N < N_{c,0}$, with
\begin{equation}
    N_{c,0} = {3\over \sqrt{2}} \sqrt{1 + \sqrt{3}} \approx 3.506.
\end{equation}
For $N = N_{c,0}$ these two fixed points merge and then, for $N>N_{c,0}$,
they become complex. For $N=3$ the stable fixed point merges with the 
O(8) fixed point.  These results show that, for integer values of $N$
satisfying $N\ge 4$, there is no stable fixed point close to four 
dimensions, hence only first-order transitions are allowed.

\subsubsection{Five-loop $\epsilon$ expansion analysis}

In order to establish the behavior of the system for $\epsilon = 1$, we 
must determine the fate of the stable fixed point that exists for 
$N<N_{c,0}$ close to four dimensions. For finite $\epsilon$, we 
expect a stable and an unstable fixed point with $v \not = 0$ up to 
$N = N_c(\epsilon)$. The two fixed points merge for $N=N_c(\epsilon)$
and become complex for $N > N_c(\epsilon)$. In order to compute
$N_c(\epsilon)$, we expand
\begin{equation}
N_c(\epsilon) = N_{c,0} + \sum_{n=1} N_{c,n} \epsilon^n,
\end{equation}
and require 
\begin{eqnarray}
&& \beta_u(u,v,N_c) = \beta_v(u,v,N_c) = 0,
\nonumber \\
&& {\rm det}\, \Omega(u,v,N_c) = 0,
\end{eqnarray}
the last equation being a consequence of the coalescence of the 
two fixed point at $N = N_c$. A straightforward calculation gives finally
\begin{eqnarray}
N_c(\epsilon) &=& 3.5063 - 0.0309 \epsilon + 0.3229 \epsilon^2 -
                1.2927 \epsilon^3 \nonumber \\ 
     && + 7.6855 \epsilon^4 + O(\epsilon^5).
\end{eqnarray}
The expansion alternates in sign, as expected for a Borel-summable series.
Resummations using the Pad\'e-Borel method appear to be stable. 
We obtain $N_c(\epsilon = 1) = 3.54(1)$ using the series to order $\epsilon^3$
and $N_c(\epsilon = 1) = 3.59(2)$ at order $\epsilon^4$ (the number in
parentheses indicates how the estimate changes by varying the resummation
parameters). Apparently, $N_c$ varies only slightly as $\epsilon$ changes from
0 to 1. In particular, this analysis predicts the absence of 
stable FPs for any integer $N\ge 4$ in three dimensions.

\subsubsection{High-order analysis in three dimensions}
\label{homsbar}

Methods based on the expansion around four dimensions allow us to find
only the 3D FPs which can be defined, by analytic continuation, close
to four dimensions.  Other FPs, which do not have a 4D counterpart,
cannot be detected.  However, the extension of this result to the
relevant $d=3$ dimension fails in some cases.  For example, this also
occurs for the Ginzburg-Landau model of superconductors, in which a
complex scalar field couples to a gauge field: although
$\epsilon$-expansion calculations do not find a stable FP
\cite{HLM-74}, thus predicting first-order transitions, numerical
analyses of 3D systems described by the Ginzburg-Landau model show
that they can also undergo continuous transitions, see, e.g.,
Ref.~\cite{MHS-02,NK-03}.  This implies the presence of a stable FP in
the 3D Ginzburg-Landau theory---in agreement with
experiments~\cite{GN-94}.  Other examples are provided by the LGW
$\Phi^4$ theories describing frustrated spin models with noncollinear
order~\cite{CPPV-04,NO-15}, the $^3$He superfluid transition from the normal
to the planar phase~\cite{DPV-04}, and the chiral transitions of the
strong interactions in the case the U(1)$_A$ anomaly effects
are suppressed~\cite{PV-13,PW-84}.

Therefore, a more conclusive analysis requires a direct study of the
3D flow. This is achieved by an alternative analysis of the
$\overline{\rm MS}$ series: the 3D $\overline{\rm MS}$ scheme without
$\epsilon$ expansion~\cite{Dohm-85,SD-89,CPPV-04}. The RG functions
$\beta_{u,v}$ are the $\overline{\rm MS}$ functions.  However,
$\epsilon\equiv 4-d$ is no longer considered as a small quantity, but
it is set equal to its physical value ($\epsilon=1$ in our case)
before computing the FPs.  This provides a well defined 3D
perturbative scheme which allows us to compute universal quantities,
without the need of expanding around $d=4$~\cite{Dohm-85,SD-89}.

We look for stable FPs of the RG flow, with a finite attraction domain
in the space of the renormalized couplings $u$ and $v$.  The RG
trajectories are determined by solving the differential equations
\begin{eqnarray}
&&-\lambda {d u\over d\lambda} = \beta_u(u(\lambda),v(\lambda)),\nonumber\\
&&-\lambda {d v\over d\lambda} = \beta_v(u(\lambda),v(\lambda)),
\label{rgfloweq}
\end{eqnarray}
where $\lambda\in [0,\infty)$, with the initial conditions
\begin{eqnarray} 
&&u(0) = v(0) = 0 ,\nonumber \\
&& \left. {d u\over d\lambda} \right|_{\lambda=0} = s\equiv {u_0\over |v_0|},\qquad
\left. {d v\over d\lambda} \right|_{\lambda=0} = \pm 1, \label{ini-rgflow}
\end{eqnarray}
where $s$ parametrizes the different RG trajectories in terms of the
bare quartic parameters, and the $\pm$ sign corresponds to the RG flows
for positive and negative values of $v_0$.  In our study of the RG flow we
only consider values of the bare couplings which satisfy
Eqs.~(\ref{stabcond1}) and (\ref{stabcond2}).  

The physically relevant results are obtained by resumming the
perturbative expansions, which are divergent but Borel summable in a
large region of the renormalized parameters.  The resummation can be
done exploiting methods that take into account their large-order
behavior, which is computed by semiclassical (hence, intrinsically
nonperturbative) instanton calculations \cite{LZ-77,ZJ-book,CPV-00}.
Relevant results for the large-order behavior of the series of the
model (\ref{ahlg}) are reported in App.~\ref{sesum}.

\subsection{The 3D MZM perturbative scheme}
\label{mzm}

In the massive zero-momentum (MZM)
scheme~\cite{Parisi-80,ZJ-book,PV-02,v-07} one performs the perturbative
expansion directly in three dimensions, in the critical region of the
disordered phase, in powers of the zero-momentum renormalized quartic
couplings.  The theory is renormalized by introducing a set of
zero-momentum conditions for the one-particle irreducible two-point
and four-point correlation functions of the 2$\times$2 matrix
field $\Psi$:
\begin{eqnarray}
&&\Gamma^{(2)}_{a_1a_2,b_1b_2}(p) = 
\left(\delta_{a_1b_2}\delta_{a_2b_1} -{1\over N}
\delta_{a_1a_2}\delta_{b_1b_2} \right)  \times
\label{ren1}  \\
&&\qquad \qquad\times
Z_\psi^{-1} \left[ m^2+p^2+O(p^4)\right],\nonumber \\
&&\Gamma^{(4)}_{a_1a_2,b_1b_2,c_1c_2,d_1d_2}(0) 
= Z_\psi^{-2} m^{4-d} \times
\label{ren2}  \\
&&
\quad \times \left(u U_{a_1a_2,b_1b_2,c_1c_2,d_1d_2} +  
v V_{a_1a_2,b_1b_2,c_1c_2,d_1d_2}\right),
\nonumber 
\end{eqnarray}
where $U,\,V$ are appropriate form factors defined so that $u\propto
u_0/m$ and $v\propto v_0/m$ at the leading tree order.  The FPs of the
theory are given by the common zeroes of the Callan-Symanzik
$\beta$-functions
\begin{equation}
\beta_u(u,v) = \left. m{\partial u\over \partial m}\right|_{u_0,v_0},\quad
\beta_v(u,v) = \left. m{\partial v\over \partial m}\right|_{u_0,v_0}.
\label{betaf}
\end{equation}
The normalization of the zero-momentum quartic variables $u, v$ is
such that their one-loop $\beta$ functions read
\begin{eqnarray}
\beta_u &=& - u  + u^2  +  { 4 N^4 + 22N^2 - 42  \over N(N^2+7)^2} uv
\label{betau0}\\
&& + {3 N^4 + 30 N^2  + 63\over N^2 (N^2+7)^2 } v^2,\nonumber \\
\beta_v &=& - v  + { 12 \over N^2+7}
uv + {2 N^4 - 4 N^2 - 126 \over N(N^2+7)^2} v^2 .
\label{betav0}
\end{eqnarray}

We compute the MZM perturbative expansions of the $\beta$ functions
and of the critical exponents up to six loops, requiring the
computation of 1428 Feynman diagrams.  The complete expansion for
$N=4$ can be found in App.~\ref{hopert}.  The large-order behaviors of
the series are reported in App.~\ref{sesum}.  The RG trajectories are
obtained by solving differential equations analogous to
Eqs.~(\ref{rgfloweq}) and (\ref{ini-rgflow}), after resumming the
$\beta$ functions as outlined in App.~\ref{sesum}.

\subsection{Results}
\label{resrgflow}

\begin{figure}[tbp]
\includegraphics*[scale=\graphicscale,angle=-90]{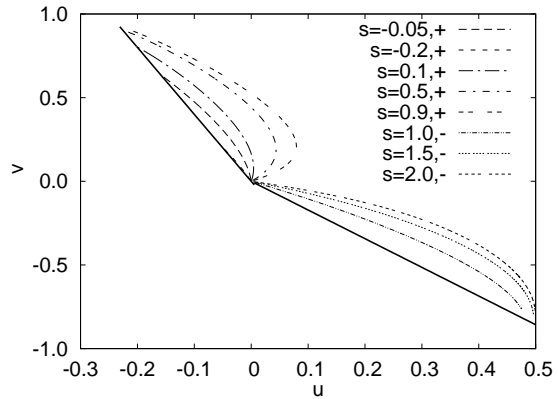}
\caption{(Color online) RG flow of the LGW theory (\ref{ahlg}) for
  $N=4$, in the ${\overline {\rm MS}}$ scheme without $\epsilon$
  expansion, for several values of the ratio $s\equiv u_0/|v_0|$ of
  the bare quartic parameters.  The curves are obtained by solving
  Eqs.~(\ref{rgfloweq}) with the initial conditions
  (\ref{ini-rgflow}): in the legend we report the value of $s$ and the
  sign of $v_0$ ("+" and "-" correspond to $v_0 > 0$ and $v_0 < 0$,
  respectively).  The two solid lines represent the boundary of the
  Borel-summability region, defined by $u + v/4 >0$ and $u + b_4 v = u
  + 7 v/12 > 0$.  }
\label{n4trams}
\end{figure}

\begin{figure}[tbp]
\includegraphics*[scale=\graphicscale,angle=-90]{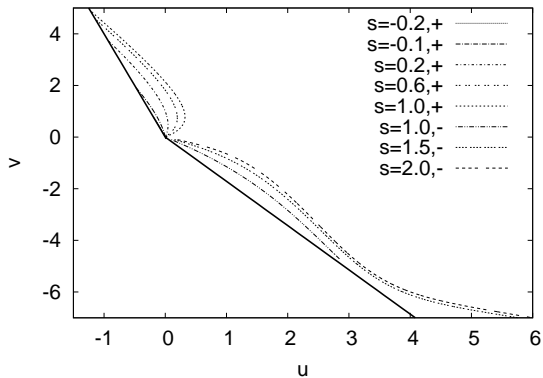}
\caption{(Color online) RG flow of the LGW theory (\ref{ahlg}) for
  $N=4$, in the MZM scheme, for several values of the ratio $s\equiv
  u_0/|v_0|$ of the bare quartic parameters.  The curves are obtained
  by solving Eqs.~(\ref{rgfloweq}) with the initial conditions
  (\ref{ini-rgflow}): in the legend we report the value of $s$ and the
  sign of $v_0$ ("+" and "-" correspond to $v_0 > 0$ and $v_0 < 0$,
  respectively).  The two solid lines represent the boundary of the
  Borel-summability region, defined by $u + v/4 >0$ and $u + b_4 v = u
  + 7 v/12 > 0$.  }
\label{n4tramzm}
\end{figure}

Some RG trajectories in the renormalized coupling space of the LGW
theory (\ref{ahlg}) for $N=4$ are shown in Figs.~\ref{n4trams} and
\ref{n4tramzm}, for the ${\overline {\rm MS}}$ and MZM schemes,
respectively, for several values of the ratio $s\equiv u_0/|v_0|$.  In
both renormalization schemes most of the RG trajectories flow towards
the region in which the series are no longer Borel summable. In the
MZM scheme, for $v_0 < 0$ some trajectories flow instead towards
infinity. In all cases, we do not have evidence of a stable FP.
Analogous results are obtained for $N=6$.

These results imply that there is no universality class characterized
by the symmetry breakings (\ref{vnpat}) and (\ref{vppat}). This would
suggest a first-order transition.  It is also possible that more than
one transition is present, each of them associated to a partial
decoupling of some degrees of freedom, hence to a different
symmetry-breaking pattern, as it happens in two-dimensional frustrated
XY models~\cite{HPV-05}. In this case, continuous transition would
still be possible.

\section{Conclusions}
\label{conclu}

We have investigated the nature of the phase transitions in 3D
ACP$^{N-1}$ models, such as the lattice model (\ref{hcpn}) with $J>0$,
which are characterized by a global U($N$) symmetry and a local U(1)
gauge symmetry.

In order to analyze their critical behavior, we construct the
corresponding LGW theory, assuming a staggered gauge-invariant order
parameter. This leads to the LGW $\Psi^4$ theory (\ref{ahlg}), where
$\Psi$ is a traceless hermitian $N\times N$ matrix, which is symmetric
under the global U($N$) transformations $\Psi \to U \Psi U^\dagger$,
and the ${\mathbb Z}_2$ transformations $\Psi \to -\Psi$.  For $N =
3$, the LGW mode is equivalent to the LGW associated with an
8-component real vector field. Hence, we predict that, if the ACP$^2$
model undergoes a continuous transition, it should belong to the O(8)
vector universality class. Note that, at the critical point, there is
an effective symmetry enlargement, U(3)$\to$O(8), and the same should
occur in the low-temperature phase as the critical point is
approached. The low-temperature symmetry ${\rm U}(1)\times {\rm U}(2)$
should be promoted to O(7).  We confirm the RG predictions by
comparing the FSS behavior of the O(8) vector and ACP$^2$ models,
obtained by MC simulations of both lattice models.  We note that the
critical behavior, characterized by the O(8) critical exponents
$\nu=0.85(2)$ and $\eta=0.0276(5)$, definitely differs from that of
ferromagnetic CP$^2$ models, for which recent
studies~\cite{NCSOS-11,KHI-11,Kaul-12,NCSOS-13} have provided
numerical evidence of continuous transitions with critical exponents
$\nu=0.536(13)$ and $\eta=0.23(2)$.

In the case of ACP$^{N-1}$ lattice models with a higher number of
components, i.e. $N\ge 4$, the identification of the order parameter
is more complex. If the order parameter is a staggered gauge-invariant
hermitian matrix as for $N=2$ and $N=3$, the associated LGW theory is
that given in Eq.~(\ref{ahlg}).  To determine the possible existence
of continuous transitions, we study the RG flow in perturbation
theory. We compute FT perturbative series in two different
renormalization schemes up to five and six loops, respectively.  The
analysis of the RG flow does not provide evidence of stable FPs.  This
implies that the dynamics of the staggered gauge-invariant modes
associated with the $N\times N$ hermitian matrices defined in
Eq.~(\ref{biab}) does not give rise to continuous transitions.  In
particular, we do not expect continuous transitions characterized by
the symmetry breaking ${\rm U}(N) \to {\rm U}(1)\times {\rm U}(N-1)$.
Thus, if the ACP$^{N-1}$ lattice model presents transitions with this
symmetry breaking, they must be first order. Note that it is still
possible to have continuous transitions if they are associated with a
different symmetry breaking, as it may arise from a partial decoupling
of some degrees of freedom.  Another possible scenario may arise from
the relevance of further gauge degrees of freedom which are not taken
into account by the LGW theory, analogously to the case of
ferromagnetic CP$^{N-1}$ model in the large-$N$ limit.  This issue
needs further investigation.

\appendix

\section{Ground state for the antiferromagnetic model with $N=3$}
\label{appmin}

\begin{figure}[tbp]
\begin{center}
\includegraphics[scale=0.8]{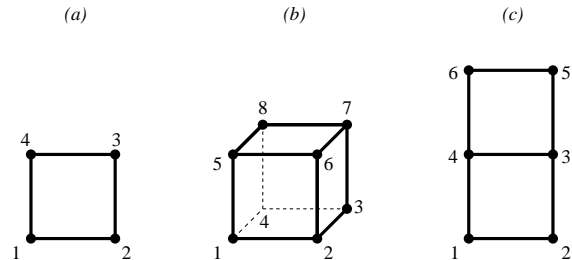}
\caption{(Color online) We draw some lattice configurations required
  by the discussion of Sec.~\ref{appmin}.  }
\label{GS_N3}
\end{center}
\end{figure}

In this section we wish to characterize the structure of the ground
state of the antiferromagnetic model for $N=3$, showing the emergence
of a ferromagnetic order on a staggered lattice. Let us first consider
a lattice plaquette, see Fig.~\ref{GS_N3}(a), and let us determine the
configurations of the four spins ${\bm z}_1$, ${\bm z}_2$, ${\bm
  z}_3$, and ${\bm z}_4$ that minimize the energy. Given the global
invariance of the model, it is not restrictive to assume that ${\bm
  z}_1 = (1,0,0)$. Since the model is antiferromagnetic, the energy is
minimized if neighbouring spins are orthogonal, i.e., if $\bar{\bm
  z}_{i}\cdot {\bm z}_j$ = 0 for nearest neughbor sites $i$ and $j$.
Therefore, we have
\begin{equation}
{\bm z}_2 = (0,{\bm v}_2), \qquad {\bm z}_4 = (0,{\bm v}_4),
\end{equation}
where ${\bm v}_2$ and ${\bm v}_4$ are two-dimensional complex unit vectors. 
If we set ${\bm z}_3 = (z_{31},{\bm v}_3)$ and consider the links
connecting site 4 with its neighbors, we obtain the conditions
\begin{equation}
\bar{\bm v}_2 \cdot {\bm v}_3 = \bar{\bm v}_4 \cdot {\bm v}_3=0.
\label{eq-v2v3v4}
\end{equation}
If ${\bm v}_2 \not= e^{i\phi} {\bm v}_4$ ($\phi$ is an arbitrary
phase), this condition implies ${\bm v}_3 = 0$. Therefore, discarding
an irrelevant phase, we obtain ${\bm z}_1 = {\bm z}_3$. If instead If
${\bm v}_2 = e^{i\phi} {\bm v}_4$, eliminating an irrelevant phase we
obtain ${\bm z}_2 = {\bm z}_4$. This analysis shows, therefore, that
in the ground-state configuration two opposite spins in the plaquette
are identical.  Note that the result holds only for $N=3$. For $N\ge
4$, Eq.~(\ref{eq-v2v3v4}) does not imply ${\bm v}_3 = 0$ in the
generic case. We only obtain that ${\bm z}_3$ belongs to
$(N-2)$-dimensional subspace containing ${\bm z}_1$.  This may leave
open the possibility of other symmetry breaking patterns for $N>3$.

Let us now assume that we are dealing with a three-dimensional system
and let us consider a cube, see Fig.~\ref{GS_N3}(b). We wish now to
show that the dominant ground-state configurations have the following
structure.  One can have ${\bm z}_1 = {\bm z}_3 = {\bm z}_6 = {\bm
  z}_8$ (we properly fix the phases), while ${\bm z}_2$, ${\bm z}_4$,
${\bm z}_5$, ${\bm z}_7$ are arbitrary spins lying in the complex
two-dimensional plane orthogonal to ${\bm z}_1$. Of course, the
equivalent arrangement with ${\bm z}_2 = {\bm z}_4 = {\bm z}_5 = {\bm
  z}_7$ is also possible.  Note that this configuration is highly
degenerate, as four two-dimensional vectors can be arbitrarily chosen.

To prove the previous statement, we consider all other possible
arrangements that are consistent with the result obtained for the
plaquette and we show that they are less degenerate that the one
discussed before.  Hence they are entropically disfavored and become
irrelevant in the infinite-volume limit. It is not restrictive to
assume that ${\bm z}_1 = {\bm z}_3 = (1,0,0)$, since two opposite
spins of plaquette $\langle 1,2,3,4 \rangle$ (we report in angular
brackets the sites belonging to the plaquette) are necessarily
identical. Let us now consider the plaquette $\langle 5,6,7,8
\rangle$.  There are three different possibilities consistent with the
result valid for a single plaquette: (i) ${\bm z}_6 = {\bm z}_8$ and
${\bm z}_6 = {\bm z}_1$; (ii) ${\bm z}_6 = {\bm z}_8$ with ${\bm z}_6
\not= e^{i\phi} {\bm z}_1$; (iii) ${\bm z}_5 = {\bm z}_7$. We wish now
to exclude cases (ii) and (iii).  In case (ii) the orthogonality
condition applied to links (1,2) and (2,6) implies
\begin{equation}
\bar{{\bm z}}_1 \cdot {\bm z}_2 = \bar{{\bm z}}_6 \cdot {\bm z}_2 = 0
\end{equation}
Since ${\bm z}_6 \not= e^{i\phi} {\bm z}_1$, these two conditions 
completely specify vector ${\bm z}_2$ (the phase is of course irrelevant).
Repeating the argument to the spins at sites 2,4,5,7 we end up with 
\begin{equation}
{\bm z}_2 = {\bm z}_4 = {\bm z}_5 = {\bm z}_7
\end{equation}
It is therefore a configuration of type (i), but now the spins are
ordered on the complementary sites 2,4,5,7.  Let us now consider case
(iii). Generically, ${\bm z}_2 \not= e^{i\phi} {\bm z}_5$ and ${\bm
  z}_4 \not= e^{i\phi} {\bm z}_5$. This implies that ${\bm z}_6$ and
${\bm z}_8$ are uniquely defined. Therefore, the configuration is
defined by specifying ${\bm z}_2$, ${\bm z}_4$, and ${\bm z}_5$, in
the two-dimensional complex space orthogonal to ${\bm z}_1$. This
configuration is less degenerate than that defined at point (i), hence
it is irrelevant in the infinite-volume limit (it is entropically
suppressed).  If ${\bm z}_2 = e^{i\phi} {\bm z}_5$ and ${\bm z}_4
\not= e^{i\phi} {\bm z}_5$, we obtain the same type of degeneracy,
since now ${\bm z}_4$, ${\bm z}_5$, and ${\bm z}_6$ can be chosen.
The other two cases give the same result.

The result for a cube extends trivially to the whole lattice, proving
that in the ground state we observe two different symmetry
breakings. First, lattice translational invariance is broken with the
emergence of a staggered symmetry. Second, on one of the two
sublattices the system orders ferromagnetically, breaking the $U(3)$
symmetry down to $U(2)$. It is important to note that the discussion
only applies in three dimensions. In two dimensions the results does
not hold. Indeed, referring to Fig.~\ref{GS_N3}(c), nothing forbids in
the two-dimensional case a configuration with ${\bm z}_1 = {\bm z}_3$,
${\bm z}_4 = {\bm z}_5$, and ${\bm z}_6\not={\bm z}_3$.

\begin{widetext}

\section{High-order field-theoretical perturbative expansions}
\label{hopert}

In this appendix we report the FT perturbative
series of the $\beta$ functions 
used in our RG analysis of Sec.~\ref{hopa}.
We only report those for $N=4$; the perturbative series for other
values of $N$ are available on request.

The five-loop $\beta$ functions of the 
${\overline{\rm MS}}$ scheme for $N=4$ are 
\begin{eqnarray}
\beta_u(u,v) &=& -\varepsilon
u+\textstyle\frac{23}{6}u^2-\frac{59}{12}u^3+\frac{24215}{1728}
u^4-\frac{2808613}{62208}u^5+\frac{2231}{19440}\pi^4
u^5+\frac{37543651}{221184}u^6-\frac{45935}{41472}\pi^4
u^6-\frac{56005}{326592}\pi^6 u^6\\ &&+\textstyle\frac{29}{12}u
v-\textstyle\frac{319}{72}u^2 v+\frac{72587}{3456} u^3
v-\frac{2556031}{31104} u^4 v+\frac{2639}{15552} \pi ^4 u^4
v+\frac{4691425207}{11943936} u^5 v-\textstyle\frac{4445671}{1866240}
\pi^4 u^5 v-\textstyle\frac{689765}{1959552} \pi^6 u^5 v\nonumber
\\ &&+\textstyle\frac{19}{32} v^2-\frac{1129}{576} u
v^2+\frac{210121}{13824} u^2 v^2-\frac{36468307}{497664}u^3
v^2+\textstyle\frac{17129}{155520} \pi^4 u^3
v^2+\frac{42566947705}{95551488} u^4
v^2-\textstyle\frac{71427821}{29859840} \pi^4 u^4 v^2\nonumber
\\ &&-\textstyle\frac{5358835}{15676416} \pi^6 u^4
v^2-\textstyle\frac{83}{192}v^3+\textstyle\frac{10789}{2048} u
v^3-\frac{35875069}{995328} u^2 v^3+\frac{27641}{622080} \pi^4 u^2
v^3+\textstyle\frac{27514775011}{95551488} u^3
v^3-\frac{21388189}{14929920}\pi^4 u^3 v^3\nonumber
\\ &&-\textstyle\frac{9436835}{47029248} \pi^6 u^3
v^3+\frac{243899}{442368} v^4-\frac{139893917}{15925248} u
v^4+\frac{64219}{4976640} \pi^4 u
v^4+\textstyle\frac{158989734779}{1528823808}u^2
v^4-\textstyle\frac{17049203}{31850496} \pi^4 u^2 v^4\nonumber
\\ &&-\textstyle\frac{55096345}{752467968} \pi^6 u^2
v^4-\frac{13810271}{15925248} v^5+\frac{38971}{19906560} \pi^4
v^5+\frac{60552906587}{3057647616}u
v^5-\textstyle\frac{6124463}{53084160} \pi^4 u
v^5-\frac{3213535}{214990848} \pi^6 u v^5\nonumber
\\ &&+\textstyle\frac{39907063243}{24461180928}
v^6-\frac{83027651}{7644119040} \pi^4
v^6-\frac{4971655}{4013162496}\pi^6 v^6+\textstyle\frac{97}{18} u^4
\zeta(3)-\frac{27967}{648} u^5
\zeta(3)+\textstyle\frac{1058293}{4608}u^6 \zeta(3)+\frac{58}{9} u^3 v
\zeta(3)\nonumber \\ &&-\textstyle\frac{400519}{5184} u^4 v
\zeta(3)+\frac{127075013}{248832} u^5 v
\zeta(3)+\textstyle\frac{203}{48} u^2 v^2 \zeta(3)-\frac{232333}{3456}
u^3 v^2\zeta(3)+\frac{371192045}{663552} u^4 v^2
\zeta(3)+\frac{83}{48} u v^3 \zeta(3)\nonumber
\\ &&-\textstyle\frac{697141}{20736} u^2 v^3
\zeta(3)+\textstyle\frac{242855651}{663552} u^3 v^3
\zeta(3)+\frac{1271}{4608} v^4 \zeta(3)-\frac{742109}{82944} u v^4
\zeta(3)+\frac{4536772733}{31850496} u^2 v^4
\zeta(3)-\frac{1246357}{1327104} v^5 \zeta(3)\nonumber
\\ &&+\textstyle\frac{211486009}{7077888} u v^5
\zeta(3)+\frac{435745159}{169869312} v^6 \zeta(3)-\frac{291}{32} u^6
\zeta(3)^2-\frac{113129}{5184} u^5 v \zeta(3)^2-\frac{887615}{41472}
u^4 v^2 \zeta(3)^2-\textstyle\frac{1485335}{124416} u^3 v^3
\zeta(3)^2\nonumber \\ &&-\textstyle\frac{8714053}{1990656} u^2 v^4
\zeta(3)^2-\frac{4430365}{3981312} u v^5
\zeta(3)^2-\frac{1655603}{10616832} v^6 \zeta(3)^2-\frac{2435}{54} u^5
\zeta(5)+\textstyle\frac{431011}{864} u^6 \zeta(5)-\frac{2755}{36} u^4
v \zeta(5)\nonumber \\ &&+\textstyle\frac{2847481}{2592} u^5 v
\zeta(5)-\frac{85975}{1296} u^3 v^2 \zeta(5)+\frac{15884369}{13824}
u^4 v^2 \zeta(5)-\textstyle\frac{43825}{1296} u^2 v^3
\zeta(5)+\frac{5504765}{7776} u^3 v^3 \zeta(5)-\frac{364135}{41472} u
v^4 \zeta(5)\nonumber \\ &&+\textstyle\frac{518304421}{1990656} u^2
v^4 \zeta(5)-\frac{11345}{13824} v^5
\zeta(5)+\textstyle\frac{104582551}{1990656} u v^5
\zeta(5)+\frac{15744751}{3538944} v^6 \zeta(5)+\frac{319039}{864} u^6
\zeta(7)+\frac{444773}{576} u^5 v \zeta(7)\nonumber
\\ &&+\textstyle\frac{3807055}{4608} u^4 v^2
\zeta(7)+\textstyle\frac{7509005}{13824} u^3 v^3
\zeta(7)+\frac{15674365}{73728} u^2 v^4 \zeta(7)+\frac{733383}{16384}
u v^5 \zeta(7)+\frac{1518167}{393216} v^6 \zeta(7),\nonumber
\end{eqnarray}
\begin{eqnarray}
\beta_v(u,v) &=& -\varepsilon v+2 u v-\textstyle\frac{157}{36} u^2
v+\frac{5879}{864} u^3 v-\frac{685387}{20736} u^4 v+\frac{19}{108}
\pi^4 u^4 v+\frac{4545155}{46656} u^5 v-\frac{989059}{933120}\pi ^4
u^5 v-\frac{24925}{122472} \pi^6 u^5 v\\ &&+\textstyle\frac{7}{12}
v^2-\frac{229}{72} u v^2+\frac{5921}{864} u^2
v^2-\frac{3207107}{62208} u^3 v^2+\frac{1073}{3888} \pi^4 u^3
v^2+\frac{2236916635}{11943936} u^4 v^2-\frac{379369}{186624} \pi^4
u^4 v^2-\frac{767575}{1959552} \pi^6 u^4 v^2\nonumber
\\ &&-\textstyle\frac{29}{64} v^3+\frac{40699}{13824} u
v^3-\frac{5568277}{165888} u^2 v^3+\frac{12247}{77760} \pi^4 u^2
v^3+\frac{2050563751}{11943936} u^3 v^3-\frac{11735557}{7464960} \pi^4
u^3 v^3-\frac{1806125}{5878656} \pi^6 u^3 v^3\nonumber
\\ &&+\textstyle\frac{5885}{9216} v^4-\frac{3433885}{331776} u
v^4+\frac{133}{3456} \pi^4 u v^4+\frac{8406050713}{95551488} u^2
v^4-\frac{381881}{622080} \pi^4 u^2 v^4-\frac{11779825}{94058496}
\pi^6 u^2 v^4-\frac{18033929}{15925248} v^5\nonumber
\\ &&+\textstyle\frac{4}{1215} \pi^4 v^5+\frac{4316967439}{191102976}
u v^5-\frac{9687473}{79626240} \pi^4 u v^5-\frac{5139325}{188116992}
\pi^6 u v^5+\frac{6609591883}{3057647616}
v^6-\frac{2367521}{238878720} \pi^4 v^6\nonumber
\\ &&-\textstyle\frac{163685}{62705664} \pi^6 v^6+\frac{58}{9} u^3 v
\zeta(3)-\frac{4987}{144} u^4 v \zeta(3)+\frac{2807371}{15552} u^5 v
\zeta(3)+\frac{43}{6} u^2 v^2 \zeta(3)-\frac{132667}{2592} u^3 v^2
\zeta(3)\nonumber \\ &&+\textstyle\frac{3169315}{9216} u^4 v^2
\zeta(3)+\textstyle\frac{175}{72} u v^3 \zeta(3)-\frac{75677}{2592}
u^2 v^3 \zeta(3)+\frac{2558749}{9216} u^3 v^3 \zeta(3)+\frac{145}{576}
v^4 \zeta(3)-\frac{326005}{41472} u v^4 \zeta(3)\nonumber
\\ &&+\textstyle\frac{235439287}{1990656} u^2 v^4
\zeta(3)-\textstyle\frac{300469}{331776} v^5
\zeta(3)+\frac{8893439}{331776} u v^5 \zeta(3)+\frac{6145307}{2359296}
v^6 \zeta(3)-\frac{1099}{324} u^5 v \zeta(3)^2-\frac{33215}{5184} u^4
v^2 \zeta(3)^2\nonumber \\ &&-\textstyle\frac{37625}{15552} u^3 v^3
\zeta(3)^2 +\textstyle\frac{364091}{248832} u^2 v^4
\zeta(3)^2+\frac{650747}{497664} u v^5 \zeta(3)^2+\frac{88027}{331776}
v^6 \zeta(3)^2-\frac{2485}{54} u^4 v \zeta(5)+\frac{540673}{1296} u^5
v \zeta(5)\nonumber \\ &&-\textstyle\frac{11035}{162} u^3 v^2
\zeta(5)+\textstyle\frac{228535}{288} u^4 v^2
\zeta(5)-\frac{50945}{1296} u^2 v^3 \zeta(5)+\frac{19449871}{31104}
u^3 v^3 \zeta(5)-\frac{59095}{5184} u v^4
\zeta(5)+\frac{4044073}{15552} u^2 v^4 \zeta(5)\nonumber
\\ &&-\textstyle\frac{7285}{4608} v^5 \zeta(5)
+\textstyle\frac{58034203}{995328} u v^5
\zeta(5)+\frac{3784553}{663552} v^6 \zeta(5)+\frac{29155}{72} u^5 v
\zeta(7)+\frac{221725}{288} u^4 v^2 \zeta(7)+\frac{354515}{576} u^3
v^3 \zeta(7)\nonumber \\ &&+\textstyle\frac{2424275}{9216} u^2 v^4
\zeta(7)+\textstyle\frac{560413}{9216} u v^5
\zeta(7)+\frac{683795}{110592} v^6 \zeta(7).\nonumber
\end{eqnarray}

The six-loop $\beta$ functions of the 
MZM scheme for $N=4$ are 
\begin{eqnarray}
\beta_u(u,v) &=& -u+ u^2-0.22544283 u^3+0.10908673 u^4-0.06576687
u^5+0.04692261 u^6-0.03823225 u^7\\ &&+0.63043478 u v-0.20303858 u^2
v+0.15749967 u^3 v-0.11634780 u^4 v+0.10631128 u^5 v-0.10058595 u^6
v\nonumber \\ &&+0.15489130 v^2-0.08970454 u v^2+0.11442230 u^2
v^2-0.10401712 u^3 v^2+0.12188821 u^4 v^2-0.13382427 u^5 v^2\nonumber
\\ &&-0.01961248 v^3+0.04186104 u v^3-0.05255068 u^2 v^3+0.08208912
u^3 v^3-0.10859850 u^4 v^3+0.00498971 v^4\nonumber \\ &&-0.01301539 u
v^4+0.03157946 u^2 v^4-0.05477669 u^3 v^4-0.00117098 v^5+0.00642308 u
v^5-0.01679641 u^2 v^5\nonumber \\ &&+0.00054799 v^6-0.00289150 u
v^6-0.00021554 v^7,\nonumber
\end{eqnarray}
\begin{eqnarray}
\beta_v(u,v) &=& -v+0.52173913 u v-0.20023805 u^2 v+0.06703734 u^3
v-0.05601881 u^4 v+0.03225406 u^5 v\\ &&-0.03176901 u^6 v+0.15217391
v^2-0.14632780 u v^2+0.06972703 u^2 v^2-0.08566328 u^3 v^2+0.06017577
u^4 v^2\nonumber \\ &&-0.07408173 u^5 v^2-0.02133655 v^3+0.02700191 u
v^3-0.05410921 u^2 v^3+0.05068529 u^3 v^3-0.07920335 u^4 v^3\nonumber
\\ &&+0.00462087 v^4-0.01692328 u v^4+0.02380079 u^2 v^4-0.04877347
u^3 v^4-0.00215837 v^5+0.00591753 u v^5\nonumber \\ &&-0.01769814 u^2
v^5+0.00059843 v^6-0.00351442 u v^6-0.00029807 v^7.\nonumber
\end{eqnarray}
\end{widetext}

\section{Summation of the pertubartive series}
\label{sesum}

Since perturbative expansions are divergent, resummation methods must
be used to obtain meaningful results.  Given a generic quantity
$S(u,v)$ with perturbative expansion $S(u,v)= \sum_{ij} c_{ij} u^i
v^j$, we consider
\begin{equation}
S(x u,x v) = \sum_k s_k(u,v) x^k,
\label{seriesx}
\end{equation}
which must be evaluated at $x=1$. The expansion (\ref{seriesx}) in
powers of $x$ is resummed by using the conformal-mapping method
\cite{ZJ-book} that exploits the knowledge of the large-order behavior
of the coefficients, generally given by
\begin{equation}
s_k(u,v) \sim k! \,[-A(u,v)]^{k}\,k^b\,\left[ 1 + O(k^{-1})\right].
\label{lobh}
\end{equation}
The quantity $A(u,v)$ is related to the singularity $t_s$ of the Borel
transform $B(t)$ that is nearest to the origin: $t_s=-1/A(u,v)$.  The
series is Borel summable for $x > 0$ if $B(t)$ does not have
singularities on the positive real axis, and, in particular, if
$A(u,v)>0$.  The large-order behavior can be determined using
semiclassical computations, based on the computations of appropriate
instanton configurations~\cite{LZ-77,ZJ-book}.  For even
$N$, these semiclassical calculations show that the expansion is Borel
summable when
\begin{equation}
u + b_N v > 0,\qquad u + \frac{1}{N}v > 0,
\label{brr}
\end{equation}
where $b_N$ is given in Eq.~(\ref{stabcond1}).  For odd $N$ we obtain 
analogously 
\begin{equation}
u + b_N v > 0,\qquad u + c_N v > 0,
\label{brr1}
\end{equation}
where $c_N$ is given in Eq.~(\ref{stabcond3}).  
Note that the
conditions for Borel summability on the renormalized couplings
correspond to the stability conditions (\ref{stabcond1}) and
(\ref{stabcond2}) of the bare quartic couplings.  In this
Borel-summability region we have for even $N$
\begin{equation}
A(u,v) = \frac{1}{2} \,{\rm Max} \left( u+b_N v,u+v/N\right).
\label{afg}
\end{equation}
For odd $N$, we should replace $u+v/N$ with $u + c_N v$.
Under the additional assumption that the Borel-transform singularities
lie only in the negative axis, the conformal-mapping method turns the
original expansion into a convergent one in the region (\ref{brr}).
Outside, the expansion is not Borel summable.

Analogously one can derive the large-order behavior of the MZM scheme,
which is again given by Eq.~(\ref{lobh}) but with
\begin{equation}
A(u,v) = {1.32997 \over N^2 + 7} \,{\rm Max} \left( u+b_N v,u+v/N\right)
\label{afgmzm}
\end{equation}
for even $N$. For odd values of $N$,
$u+v/N$ should be replaced with $u + c_N v$.

We use the conformal mapping method to resum the series taking into
account what we know about their large-order behavior.  The method we
use is described in Refs.~\cite{ZJ-book,CPV-00}.  Resummations depend
on two parameters, which are optimized in the procedure.  The
approximants we use depend on two parameters $\alpha$ and $b$; we use
here the notations of Refs.~\cite{ZJ-book,CPV-00}.

\end{document}